\documentclass[fleqn,11pt]{article}
\usepackage{graphicx,array}
\usepackage{amssymb,amsmath}
\usepackage{cite}
\usepackage{bm}
\usepackage[usenames]{color}
\renewcommand{\normalsize}{\fontsize{9pt}{15pt}\selectfont}
\newcommand{\halfnormalsize}{\fontsize{9pt}{12pt}\selectfont}
\renewcommand\title[1]{
\begin{center}\fontsize{12pt}{15pt}{\bf{\bfseries\sffamily#1}\par}\end{center}}
\renewcommand\author[1]{\begin{center}\small{\sffamily#1}\end{center}\par}
\renewcommand\abstract[1]{\begin{quote}\footnotesize #1\end{quote}\par}
\makeatletter
\renewcommand{\fnum@figure}[1]{\fontsize{8pt}{12pt}{\textbf{Figure~\thefigure. }}}
\renewcommand{\fnum@table}[1]{\fontsize{8pt}{12pt}{\textbf{Table~\thetable. }}}
\@ifundefined{captionstyle}{}{}
\@ifundefined{instindent}{\newdimen\instindent}{}
\long\def\@caption#1[#2]#3{\par\addcontentsline{\csname
  ext@#1\endcsname}{#1}{\protect\numberline{\csname
  the#1\endcsname}{\ignorespaces #2}}\begingroup
    \@parboxrestore\if@minipage\@setminipage\fi
    \@makecaption{\csname fnum@#1\endcsname}{\ignorespaces #3}\par
  \endgroup}
\renewcommand\@seccntformat[1]{\csname prefmt@#1\endcsname
	\csname the#1\endcsname. \csname postfmt@#1\endcsname}
\newcommand\postfmt@section{\hskip 1mm}
\newcommand\postfmt@subsection{\hskip 1mm}
\renewcommand\section{\@startsection{section}{1}{\z@}%
{-12pt plus3pt minus2pt}%
{3pt}%
	{\normalfont\normalsize\bfseries}}
\renewcommand\subsection{\@startsection{subsection}{2}{\z@}%
{-9pt plus3pt minus2pt}%
{3pt}%
	{\normalfont\normalsize\bfseries}}
\renewcommand\subsubsection{\@startsection{subsubsection}{3}{\z@}%
{-9pt plus3pt minus2pt}%
{3pt}%
	{\normalfont\normalsize\bfseries}}
\makeatother
\newcommand\stamp[6]{\vspace{0.25cm}\linespread{1}\noindent\footnotesize
#1,\ #2:\ #3,\ #4 ({\em #5}).\ #6\par}
\def\s{\sigma}

\def\sgn{\,\text{\rm sgn}\,}
\def\({\left(}
\def\){\right)}

\newcommand{\eq}[2]{\begin{equation}\label{#1}\begin{split}#2\end{split}\end{equation}}

\newcommand{\vv}[1]{\boldsymbol #1}					
\newcommand{\MM}[1]{\mathsf #1}						

\newcommand{\bert}{\raise-0.45mm\hbox{\large$\Box$}\,}		

\definecolor{BrickRed}{cmyk}{0,0.89,0.94,0.28}				
\definecolor{MidnightBlue}{cmyk}{0.98,0.13,0,0.43}			
\definecolor{DarkGreen}{rgb}{0.100806,0.495968,0.209979}
\definecolor{orange}{rgb}{0.587167,0.354498,0.146197}
\begin{document}
\vspace*{5.4 mm}
%
%
\title{Irreversibility of mechanical and hydrodynamic instabilities}
\vspace*{3.8 mm}
%
%
\author{Carlos D.~D\'iaz-Mar\'in, Alejandro Jenkins}
%
%
\abstract{{\em Abstract:} The literature on dynamical systems has, for the most part, considered self-oscillators (i.e., systems capable of generating and maintaining a periodic motion at the expense of an external energy source with no corresponding periodicity) either as applications of the concepts of limit cycle and Hopf bifurcation in the theory of differential equations, or else as instability problems in feedback control systems.  Here we outline a complementary approach, based on physical considerations of work extraction and thermodynamic irreversibility.  We illustrate the power of this method with two concrete examples: the mechanical instability of rotors that spin at super-critical speeds, and the hydrodynamic Kelvin-Helmholtz instability of the interface between fluid layers with different tangential velocities.  Our treatment clarifies the necessary role of frictional or viscous dissipation (and therefore of irreversibility), while revealing an underlying unity to the physics of many irreversible processes that generate mechanical work and an autonomous temporal structure (periodic, quasi-periodic, or chaotic) in the presence of an out-of-equilibrium background.}

%
%
\section{Introduction}\label{sec:intro}

A self-oscillator is a physical system that excites and maintains a periodic variation at the expense of a source of energy lacking any corresponding periodicity.  This definition is due to mathematical physicist A.~A.~Andronov (1901--1952) and his school, but the same class of phenomena are referred to by many other names and its scientific study dates back to the work of mechanical engineer Robert Willis (1800--1875) and mathematical astronomer Sir George Airy (1801--1892) on the operation of the vocal cords; see \cite{SO} and references therein.  Self-oscillators are described by homogeneous equations of motion, distinguishing them from forced and parametric resonators.  The treatment of self-oscillators in the scientific and engineering literatures has been largely based on the concepts of limit cycles and Hopf bifurcations in the theory of differential equations, or of instability in the theory of feedback control systems.

This work is part of an effort to develop and promote a more physical perspective on self-oscillators, based on considerations of energy, work, and efficiency.  This effort is inspired by the observation made long ago by applied physicist Philippe Le Corbeiller (1891--1980) that cyclical motors are self-oscillators, so that the study of self-oscillators may benefit from the thermodynamic perspective and vice-versa \cite{LeC1, LeC2}.  Unfortunately, Le Corbeiller did not develop this idea very far and it was not taken up by others.

We begin in Sec.~\ref{sec:rotors} by reviewing and generalizing an analysis, due to Shen and Mote \cite{Mote}, of how a rotating dashpot can transfer some of its mechanical energy into the oscillation of the elastic disk over which it moves, thus causing the disk to self-oscillate transversely when the dashpot's speed of rotation exceeds the oscillation's phase velocity.  This analysis is instructive because it reveals how the process depends on dissipation within the dashpot.  We point out that a similar analysis applies to a large class of instabilities in mechanical rotors, including the well known problem of ``shaft whirling'' in mechanical engineering.

In Sec.~\ref{sec:KH} we consider the hydrodynamical Kelvin-Helmholtz (KH) instability, by which, e.g., the action of a steady wind makes waves on the surface of a body of water.  We point out the close analogy between this instability and the mechanical ones considered previously.  We then review the simple argument used by theoretical physicist Y.~B.~Zel'dovich (1914 -- 1987) to deduce, from the same considerations that account for the KH instability, that a spinning black hole should radiate \cite{Zeldovich, Zeldovich2, Bekenstein, Superradiance}.  This illustrates the power of thermodynamic reasoning to abstract and generalize across diverse physical phenomena.

In Sec.~\ref{sec:dissipate} we discuss how these various phenomena illuminate the physics of what applied mathematician Jerry Marsden (1942 -- 2010) and collaborators called ``dissipation-induced instabilities''; see \cite{Marsden} and references thererin.  We shall see that, far from being a paradoxical curiosity, the fact that dissipative forces may destabilize an equilibrium reflects the elementary facts that cyclical engines can be powered only by non-conservative forces and that, according to the second law of thermodynamics, any non-conservative force must be accompanied by the generation of entropy (i.e., by dissipation).  Thus, the ubiquity of self-oscillators (from turbines to neurons) reflects the {\it thermodynamic irreversibility} of macroscopic physical interactions.  We conclude by connecting this to the observation, stressed in the recent literature on ``finite-time thermodynamics'', that a cyclical engine capable of delivering non-zero power must operate irreversibly (see \cite{Ouerdane} and references therein).

%
%
\section{Rotors}\label{sec:rotors}

In \cite{Mote}, Shen and Mote studied the possible mechanisms of instability of an elastic disk under a rotating spring-mass-dashpot system, and found that a viscous dashpot destabilizes the disk when the dashpot moves faster than the phase velocity of a transverse wave on the free disk.  Here we review their result, considering a general dissipative force of the dashpot on the disk and then underlining how this analysis can be generalized to other systems.

%
\subsection{Action of rotating dashpot on elastic disk}\label{sec:Mote}

Let $w(t, r, \theta)$ be the transverse displacement at time $t$ of a mass element of the disk corresponding to polar coordinates $(r, \theta)$ on the disk's equilibrium plane.  We work in a frame of reference in which the disk does not rotate and write $w_t \equiv \partial w / \partial t$, $w_\theta \equiv \partial w / \partial \theta$.  The dashpot moves with angular velocity $\dot \theta \equiv d \theta / dt$ and $r = r_0 = $ const.  The transverse velocity of the disk element in contact with the dashpot is $\dot w \equiv d w / dt = w_t + \dot \theta w_\theta$.  The dashpot exerts a dissipative force that resists this transverse displacement:
\eq{eq:Fdis}{
F_{\rm dis} = -\sgn{(\dot w)} f_{\rm pos}}
where $f_{\rm pos}$ is arbitrary but strictly non-negative (Shen and Mote take $f_{\rm pos} = c |\dot w|$ for a constant $c > 0$, corresponding to linear damping).

\begin{figure}[t]
\center
\includegraphics[scale=0.45]{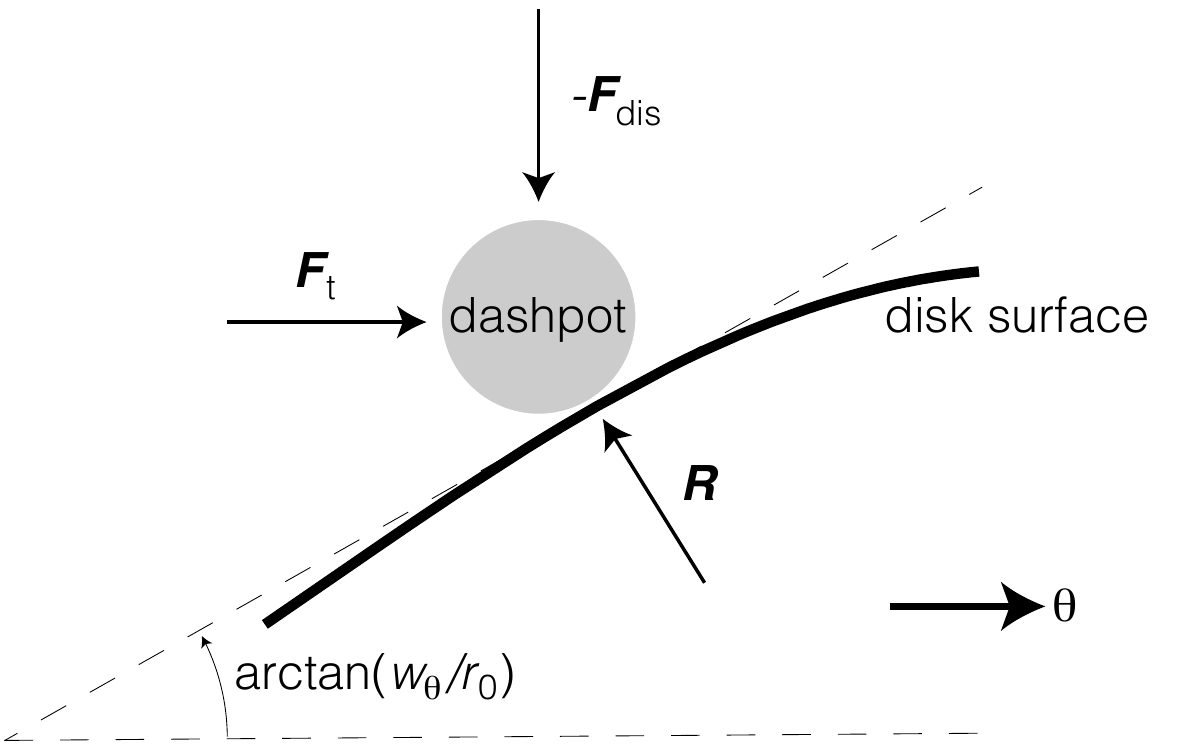}
\caption{Free-body diagram for a dashpot in contact with the surface of an elastic disk.}
\label{fig:Mote}
\end{figure}

Taking the dashpot to be massless (or, equivalently, requiring that its kinetic energy remain constant) and neglecting the friction between the disk and the dashpot, the tangential force $F_{\rm t}$ required to keep the dashpot in uniform circular motion with $\dot \theta = \Omega =$ const.\ is
\eq{eq:Ft}{
F_{\rm t} = - \left[ \frac{F_{\rm dis} \cdot w_\theta}{r_0} \right]_{\theta = \Omega t, \, r=r_0} =
\left[ \frac{\sgn{(\dot w)} f_{\rm pos} w_\theta}{r_0} \right]_{\theta = \Omega t, \, r=r_0}}
(see Fig.~\ref{fig:Mote}).  The work done by this force over a period $\tau$ is
\eq{ec:Wt}{
W_{\rm t} = \int_0^\tau F_{\rm t} r_0 \Omega dt =
\Omega \int_0^\tau \left[ \sgn{(\dot w)} f_{\rm pos} w_\theta \right]_{\theta = \Omega t, \, r=r_0} dt}
and the energy dissipated in the dashpot is
\eq{eq:Wd}{
W_{\rm d} = - \int_0^\tau \left[ F_{\rm dis} \dot w \right]_{\theta = \Omega t, \, r=r_0} dt = \int_0^\tau \left[ \sgn{(\dot w)} f_{\rm pos} \dot w \right]_{\theta = \Omega t, \, r=r_0} dt \geq 0.}
The energy absorbed by the oscillation is therefore
\eq{eq:DE1}{
\Delta E = W_{\rm t} - W_{\rm d} &=
\int_0^\tau \left[ \left( \Omega w_\theta - \dot w \right) \sgn{(\dot w)} f_{\rm pos} \right]_{\theta =
\Omega t, \, r=r_0} dt \\
&= - \int_0^\tau \left[ w_t \sgn{(\dot w)} f_{\rm pos} \right]_{\theta = \Omega t, \, r=r_0} dt
= \int_0^\tau \left[ w_t F_{\rm dis} \right]_{\theta = \Omega t, \, r=r_0} dt. }
If we consider a traveling wave of the form $w = A(r) \sin \left( m \theta - \omega t \right)$ for $\omega \geq 0$ and define a parameter $\s \equiv m \Omega - \omega$ this becomes
\eq{eq:DE2}{
\Delta E &= \int_0^\tau \left[ \omega A(r) \cos \left( m \theta - \omega t \right) \sgn{(\dot w)} f_{\rm pos}\right]_{\theta = \Omega t, \, r=r_0} dt \\
&= \omega A(r_0) \int_0^\tau \cos(\s t) \sgn{ \left[ A(r_0) \s \cos(\s t) \right]} \left. f_{\rm pos} \right|_{\theta = \Omega t, \, r=r_0} dt \\
& = \sgn{(\s)} \cdot \omega |A(r_0)| \int_0^\tau |\cos (\s t)| \left. f_{\rm pos} \right|_{\theta = \Omega t, \, r=r_0} dt
.}
The last integral in Eq.~\eqref{eq:DE2} is strictly non-negative.  If $f_{\rm pos} \neq 0$, then \hbox{$\sgn{(\Delta E)}=\sgn{(\s)}$}.  This means that if the dashpot moves with $\Omega$ less than the phase velocity $\omega / m$, then $\Delta E < 0$, indicating that the dashpot damps the transverse oscillation of the disk.  We call this the sub-critical regime.  On the other hand, when the dashpot moves with $\Omega$ greater than the wave's phase velocity, $\Delta E > 0$, which means that the transverse oscillation is powered by the dashpot's motion.  We call this the super-critical regime.

One way of understanding the change of sign of $\Delta E$ is to note that, according to Eq.~\eqref{eq:DE1}, the power delivered to the oscillation is $w_t F_{\rm dis}$, where $w_t$ is measured with respect to the static disk's equilibrium position.  When $\Omega < \omega / m$, the force $F_{\rm dis}$ {\it lags} behind the oscillation, because of dissipation in the dashpot.  The corresponding work done on the oscillation is therefore negative.  When $\Omega > \omega / m$ the oscillation travels backwards with respect the dashpot, so that $F_{\rm dis}$ {\it leads} the oscillation, making the work on it positive. \cite{Mote,Crandall}

This analysis is more generalizable than it might seem at first.  For starters, rather than a massless dashpot maintained at constant angular velocity by an external force $F_{\rm t}$, one could take $W_{\rm t}$ as coming out of a massive dashpot's kinetic energy, causing it to decelerate. If the dashpot were initially moving super-critically ($\s > 0$), this would power the self-oscillation until the dashpot's velocity fell below the critical $\Omega = \omega / m$.  This is also equivalent to considering the dashpot to be at rest and endowing the spinning disk with kinetic energy.  A similar result is obtained for the stability of a circular saw subject to an in-plane edge load, as in a sawmill: see \cite{Hutton} and references therein.

%
\subsection{Shaft whirling}\label{sec:whirl}

The theoretical analysis of shaft whirling dates back to the work of Kimball in 1920s, in which he argued that the whirling of a super-critically spinning shaft is an instability induced by the shaft's internal friction \cite{Kimball, Kimball2}.  In Kimball's model, the stretching or compression of the material fibers in the shaft is opposed by a dissipative force.  When the shaft turns faster than the natural frequency of the whirling, the force on the fibers {\it injects} energy into the whirling.  This destabilizes the system, much like the disk of Sec.~\ref{sec:Mote} is destabilized by the super-critically spinning dashpot.

\begin{figure}[t]
\center
\includegraphics[scale=0.45]{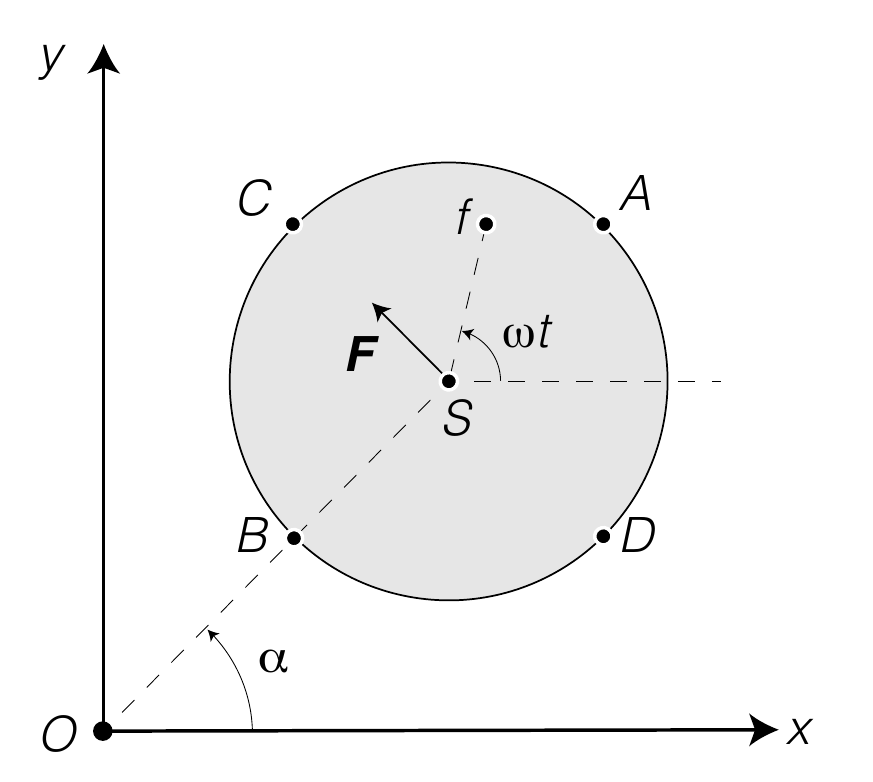}
\caption{Cross-section of a shaft turning at rate $\omega$ and whirling at rate $\dot \alpha$.  The direction of the non-conservative force $\vv F$ shown corresponds to the super-critical case $\omega > \dot \alpha$.}
\label{fig:Kimball}
\end{figure}

Consider a cross section of a shaft, centered at $S$.  A given material fiber that runs along the shaft's length passes through this cross section at a point $f$, which turns around $S$ with angular velocity $\omega$ (see Fig.~\ref{fig:Kimball}).  If the shaft is initially perturbed, displacing $S$ away from the position $O$ that it would occupy if the shaft were straight, then the shaft whirls with rate $\dot \alpha$ and amplitude $OS$.  The whirling rate $\dot \alpha$ is given by the elastic force on the shaft, which points from $S$ to $O$.  Being conservative, this restoring force does no net work over a complete period of the shaft's motion.   This may also be seen from the fact that the force always points at right angles to the whirl component of $S$'s velocity.

Following Kimball, we consider an internal friction that opposes the time rate of change of each fiber's length.  Just like the restoring force acting on $S$ points from the longer to the shorter elastic fibers, this internal friction gives rise to a force that points from the fibers being stretched to those being compressed.  This force can have a tangential component and thus do net work on the whirling shaft.  If $\omega = \dot \alpha$ then individual material fibers maintain fixed lengths, with the fiber at $B$ being shortest and the fiber at $A$ longest.  In this case the shaft experiences no internal friction and only the elastic force along $OS$ acts on the shaft.  But if $\omega \neq \dot \alpha$, then a fiber moving from $A$ to $B$ is being shortened, while a fiber moving from $B$ to $A$ is being stretched.

If the shaft turns sub-critically ($\omega < \dot \alpha$) for positive $\dot \alpha$ and $\omega$, then the fiber at $C$ is under frictional tension and the fiber at $D$ under frictional compression, and the resulting force $\vv F$ points against the whirl, damping its amplitude $OS$.  If the sign of $\dot \alpha$ is flipped then so is the sign of $\vv F$, so that whirling in either direction is damped.  On the other hand, if the shaft turns super-critically ($\omega > \dot \alpha$) the fiber at $C$ is under frictional compression and the fiber at $D$ is under frictional tension, regardless of the sign of $\dot \alpha$.  The force $\vv F$ will therefore inject energy into a whirl with $\dot \alpha > 0$, destabilizing the rotating shaft's straight configuration.

%
\subsection{Non-conservative positional force}\label{sec:NPF}

In the super-critical case, the equations of motion for the rectangular coordinates $(x,y)$ of the shaft's center of mass at $S$ take the form
\eq{eq:NPF}{ 
\left\{ \begin{array}{l}
m \ddot x + k x + p y = 0 \\
m \ddot y + k y - p x =0 \end{array} \right.}
where $k$ is the elastic constant for the bending of the fibers and $p$ is equal to the magnitude of the tangential force $F$ divided by the radius $OS$.  The terms with $p$ in Eq.~\eqref{eq:NPF} correspond, in the language of \cite{Marsden}, to a ``non-conservative positional force'' (NPF).  As a vector \hbox{${\vv F} = (-py, px, 0)$} has non-zero circulation \hbox{${\bm \nabla} \times {\vv F} = (0, 0, 2p)$} and is therefore not expressible as $- \bm \nabla V$ for any potential $V$.

\begin{figure}[t]
\center
\includegraphics[scale=0.25]{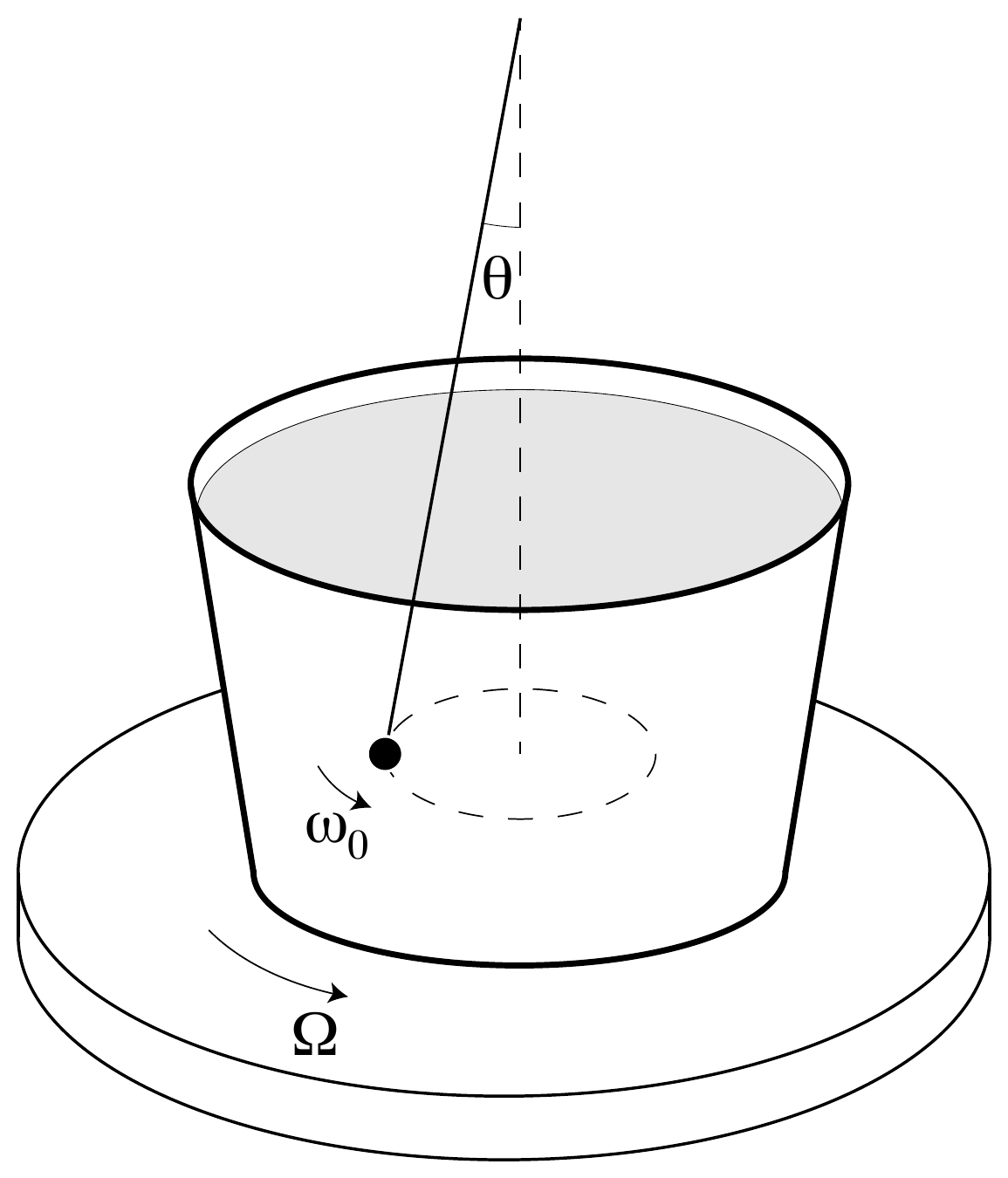}
\caption{Conical pendulum moving with amplitude $\theta$ and angular velocity $\omega_0$ in a bucket of water that spins at rate $\Omega$.  Image adapted from \cite{Crandall}.}
\label{fig:bucket}
\end{figure}

A simple analogy, originally due to physicist Sir Brian Pippard (1920--2008), helps clarify the origin of the NPF: Consider the conical pendulum swinging inside a rotating bucket filled with water, shown in Fig.~\ref{fig:bucket}.  If the water rotates more slowly than the free pendulum, then the water's viscosity damps the pendulum's motion, causing it to sink towards the vertical ($\theta \to 0$).  If the water rotates faster than the free pendulum, the water drags the pendulum forwards, causing the amplitude $\theta$ to increase \cite{Pippard,Crandall}.  Only in the latter case can the water's effect on the pendulum be described by an NPF.  Note that a model in which the NPF is obtained without reference to the elastic force that determines the critical speed is therefore inconsistent; cf.\ \cite{Merkin, Crandall2}.

%
%
\section{Hydrodynamic instabilities}\label{sec:KH}

\begin{figure}[t]
\center
\includegraphics[scale=0.4]{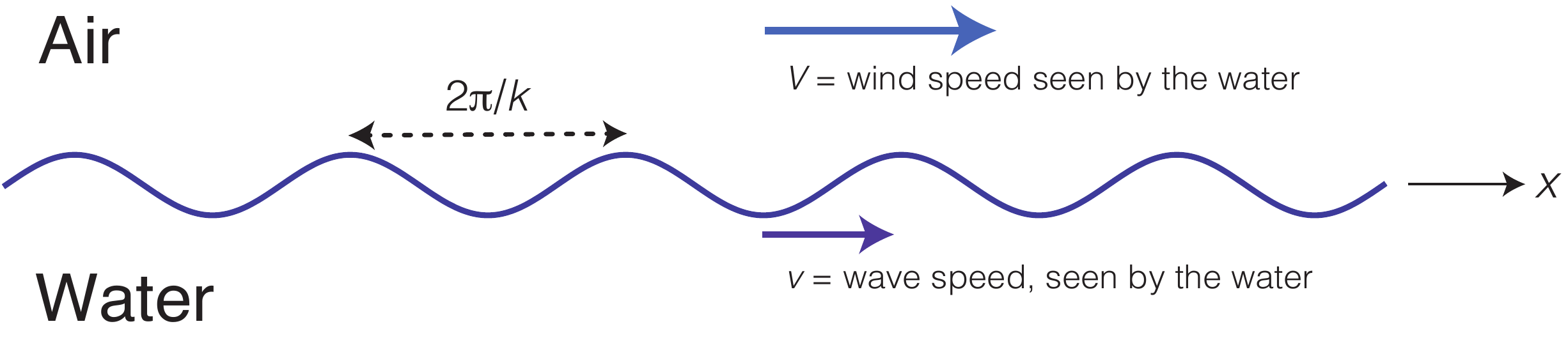}
\caption{Illustration of the hydrodynamic KH instability by which wind can generate waves on the surface of a body of water.  Image adapted from \cite{Kip-talk}.}
\label{fig:KH}
\end{figure}

The interface between two layers of fluid with different tangential velocities is unstable against a traveling transverse perturbation when the difference in the velocities of the layers exceeds the phase velocity of the perturbation with respect to the fluid.  This is known as the Kelvin-Helmholtz (KH) instability (see, e.g., \cite{Tritton}) and it is the fundamental mechanism by which a steady wind makes waves on the surface of a body of water, as illustrated in Fig.~\ref{fig:KH}.  Since the wind has no periodicity corresponding to the water wave, this is a self-oscillation.

Neglecting the viscosity of the water, in the linear regime a wave that propagates in the $x$ direction along the surface of the water and that has wave number $k$ can be expressed as the real part of
\eq{eq:free}{
\xi_0 = A \cdot e^{i(kx - \omega t)}, ~\hbox{for}~ \omega = k v.}
If we take into account the air's viscosity, then when the bulk of the air is a rest ($V=0$ in Fig.~\ref{fig:KH}) the water wave $\xi$ is described by an equation of motion that includes a linear damping term of the form $c \cdot \partial \xi / \partial t$.  If the air blows with constant velocity $V > 0$, we can go to the air's rest frame by a Galilean coordinate transformation.  The damping term acting on the free wave $\xi = \xi_0$ transforms as
\eq{eq:Galileo}{
\frac{\partial \xi_0}{\partial t} \to \frac{\partial \xi_0}{\partial t} + V \frac{\partial \xi_0}{\partial x}
= -i \omega \xi_0 + V \cdot i k \xi_0 = -i \omega \xi_0 \left( 1 - \frac V v \right).}
The sign of Eq.~\eqref{eq:Galileo} flips when $V$ passes the critical value $V = v$.  The wind therefore anti-damps a water wave with phase velocity $v$ less than the wind speed $V$.

Much like in the analysis of Sec.~\ref{sec:Mote}, this reflects that fact that when $V < v$ the oscillation of the air pressure induced by the wave lags behind the wave, because of the air's viscosity, but when $V > v$ the wave travels backwards with respect the air, so that the oscillation of the air pressure leads the wave.  Without dissipation, the phase between the air pressure and the wave must be either 0 or $\pi$, preventing the air from doing any work at all on a full period of the wave.  It is therefore clear that the KH instability requires non-zero dissipation in the air, just like we saw in Sec.~\ref{sec:rotors} that some energy had to be dissipated in the dashpot for the self-oscillation of the elastic disk to be excited, and that Kimball's explanation of shaft whirling depended on internal friction.

Aeronautical engineer Erik Mollo-Christensen (1923 -- 2009) stressed that the KH instability depends on non-vanishing viscosity in \cite{Mollo}, but many leading textbooks do not make this quite so clear.  For instance, Landau and Lifshitz derive an instability for inviscid flow because they work in a limit in which the power injected into the wave is vanishing \cite{Landau}.  The same is true of Rayleigh's inviscid instability argument for rotating Couette flow \cite{Tritton}.  Indeed, a simplified account of Rayleigh's argument (see, e.g., \cite{Feynman-TC}) could lead a novice to conclude that the fluid circulation in a Taylor cell is driven by centrifugal force, which is impossible because the centrifugal force is conservative.

The simple argument of Eq.~\eqref{eq:Galileo} led Zel'dovich to conclude that any body that, when stationary, damps an incident wave must also, if its surface moves faster than the wave's phase velocity, amplify the wave at the expense of the body's kinetic energy \cite{Zeldovich, Zeldovich2, Bekenstein}.  Moreover, some of the kinetic energy lost must heat the body, making the process thermodynamically irreversible.  In this way, Zel'dovich argued in 1971 that a spinning black hole should radiate, a result that motivated the rise of black-hole thermodynamics as an active field of research in theoretical physics (for an entertaining historical account of Zel'dovich's argument and its impact, see ch.\ 12 in \cite{Kip-book}).  The process predicted by Zel'dovich is now called ``superradiance''; see \cite{Superradiance} and references therein.

%
%
\section{Dissipation-induced instabilities}\label{sec:dissipate}

Following Krechetnikov and Marsden \cite{Marsden} we may classify the possible terms in the linearized, homogeneous equation of motion for an $n$-dimensional system with an equilibrium at $\vv q = 0$ as:
\eq{eq:linear}{
\underset{\rm inertia}{\MM M \ddot{\vv q}}  + \underset{\rm dissip.}{\MM C \dot{\vv q}} + \underset{\rm gyroscopic}{\MM G \dot{\vv q}} + \underset{\rm potential}{\MM K \vv q} + \underset{\rm non-cons.}{\MM N \vv q} = 0
}
where $\MM M, \MM C$, and $\MM K$ are $n \times n$ symmetric matrices, while $\MM G$ and $\MM N$ are anti-symmetric.  The system is trivially unstable when $\MM K$ has negative eigenvalues, which corresponds to perturbing about a configuration that is not a local minimum of potential energy.  Such a system may be stabilized by $\MM G$, a phenomenon familiar from the sleeping top.  Any non-zero, positive dissipation, no matter how small, will destabilize it again: a top which is initially sleeping will always end up falling if it dissipates mechanical energy.  This is therefore a ``dissipation-induced instability''.

Self-oscillation is seen either as a negative eigenvalue of $\MM C$ or as $\MM N \neq 0$.  In a physically realistic description, negative damping results from a positive feedback involving a non-conservative dynamic not explicitly included in Eq.~\eqref{eq:linear}; see \cite{SO}.  Similarly, we have seen here that $\MM N = 0$ requires super-critical motion in a dissipative medium.  The laws of thermodynamics thus reveal something that cannot be deduced from the mathematics of Eq.~\eqref{eq:linear}: that self-oscillation is always irreversible.

Mechanical engineer Hans Ziegler (1910--1985) discovered in 1952 that some configurations of a double pendulum with a follower load could be destabilized by arbitrarily small friction at the joints.  This ``Ziegler paradox'' launched a research program on the mathematical characterization of dissipation-induced instabilities; see \cite{Verhulst} and references therein.  Ziegler's result is not so paradoxical in light of the physical approach that we advocate here: damping is always potentially destabilizing when, as Pippard put it, a moving part ``carries its dissipative mechanisms around with it'' \cite{Pippard}; see also \cite{Crandall2}.  For instance, in Fig.~\ref{fig:KH} the energy that excites the waves comes from the wind's motion.  Instability therefore depends on the air's non-zero viscosity.  The water's viscosity, on the other hand, always damps the waves and therefore can only be stabilizing.

%
\subsection{Tidal acceleration and other analogs}\label{sec:NPF}

This approach reveals interesting analogies.  For instance in Fig.~\ref{fig:Kimball} the shaft could be replaced by the Earth and $O$ by the position of the Moon.  The viscous damping of the motion of the Earth's tidal bulge acts as internal friction.  Since
\eq{eq:tidal}{
\omega = \frac{2 \pi}{1~\hbox{day}} > \dot \alpha = \frac{2 \pi}{1~\hbox{month}},
}
the Earth spins super-critically.  The net gravitational force exerted by the Moon on the tidally deformed Earth therefore has a tangential component along the Earth's orbital velocity, explaining why the semi-major axis of the Earth-Moon orbit is currently increasing by about 4 cm/yr, a phenomenon known as ``tidal acceleration''.  \cite{Hut, Superradiance}

All of the processes that we have discussed depend on dissipation within the medium whose kinetic energy powers the self-oscillation.  The non-conservative force that drives the self-oscillation is usually exerted by that same energetic medium.  But in tidal acceleration the energy comes from the Earth's rotation, while the non-conservative force comes from the Moon's gravitational field.  This is possible because the Moon's gravity acts on a spinning Earth that is being periodically deformed by the combination of tidal and viscous forces.  This gives rise to a tangential component of the Moon's tidal force, acting with respect to the Earth's center of mass as a NPF \cite{Hut}.  Something similar is seen when a child enjoys a playground swing: the force that drives the swinging is the tension of the chain that holds up the swing.  The child provides the energy by periodically deforming her body in a way that causes the chain's tension to have a component along the velocity of the child's center of mass.  That deformation results from (very complex!) irreversible processes within the child.

Other dissipation-induced instabilities include sonic booms, Kelvin wakes, and \v Cerenkov radiation.  One of us has recently argued that the Earth's Chandler wobble should also be understood in these terms: as a destabilization of the Earth's axis of rotation, powered by the circulation of geophysical fluids and associated with viscous dissipation within them. \cite{Chandler}  

%
\subsection{Towards a physical theory of engines}\label{sec:NPF}

Thermodynamic theory allows for reversible engines, such as the cycle described by military engineer Sadi Carnot (1796--1832).  Carnot realized that reversibility required the heat flow between the working substance and the external reservoirs to occur isothermally, making it infinitely slow.  The phenomenological thermodynamics that Clausius, Kelvin and others built upon Carnot's work never considers the time dependence of the state variables, giving it a qualitatively different character to that of a mechanical description.  More recent work on ``finite-time thermodynamics'' has established that obtaining non-zero power necessarily reduces the limit efficiency of a heat engine, compared to the zero-power Carnot cycle running between the same reservoirs. \cite{Ouerdane}

Conventional thermodynamics says that all of the instabilities that we have considered can have efficiencies arbitrarily close to 1, since their energy source is mechanical rather than thermal.  We have seen, however, that these processes are {\it necessarily} irreversible, since they must be accompanied by non-zero dissipation in the body that cedes energy.  A realistic, physical description of engines capable of delivering non-zero power should take into account this distinction between necessary and avoidable dissipation.

These considerations apply not just to self-oscillators with regular limit cycles, but also to any autonomous mechanical system (i.e., one describable by homogenous differential equations) that generates a temporal structure ---whether periodic, quasi-periodic, or chaotic--- as the result of non-conservative forces.  The laws of thermodynamics imply that such systems must generate entropy; see \cite{Cross} and references therein.

%
%
\section{Conclusions}\label{sec:conclusions}

Krechetnikov's and Marsden's observation that ``ubiquitous dissipation is one of the paramount mechanisms by which instabilities develop in nature'' \cite{Marsden} seems to us both broader and deeper than the authors may have realized: almost all of what makes our everyday experience of the physical world interesting depends on irreversibility.  We have illustrated this by considering several mechanical and hydrodynamic instabilities from a different perspective than the one commonly adopted in the dynamical systems literature.  This approach clarifies the role of dissipation, without entering into the details of the processes behind it.  In the case of the KH instability, for instance, all that we needed was the fact that when the bulk of the air is at rest above the water the air tends to damp out the waves on the water's surface, as one may easily verify experimentally.  From this and from some very general symmetry principles, Zel'dovich arrived at a result that applies to ocean waves just as much as to quantum fields incident on spinning black holes.

%
%
\section*{Acknowledgments}
This work was supported by the Vice-rectorate for Research of the U.\ of Costa Rica (project no.\ 112-B6-509).  AJ thanks Robert Alicki, Stan Hutton, John McGreevy, Huajiang Ouyang, Juan Sabuco, and Miguel Sanju\'an for stimulating discussions, and Bob Jaffe for help procuring Refs.~\cite{Kimball, LeC2}.  AJ was also supported by the European Union's Horizon 2020 research and innovation program under the Marie Sk{\l}odowska-Curie grant agreement no.\ 690575.  

%
%
\halfnormalsize

%
%
%

\stamp{Carlos D.~D\'iaz-Mar\'in}{B.A.}{University of Costa Rica, School of Mechanical Engineering}{36-2060, COSTA RICA}{cdiaz95@hotmail.es}{}

\stamp{Alejandro Jenkins}{Professor}{University of Costa Rica, School of Physics}{11501-2060, COSTA
RICA}{alejandro.jenkins@ucr.ac.cr}{}
%
%
\end{document}